

  \def\sa{\vskip 0.5 true cm}
  \def\sb{\vskip 1   true cm}

\magnification\magstep1
\baselineskip = 0.75 true cm
\pageno=0
\vsize = 22 true cm
\hsize = 16.8 true cm
\parskip = 0.50 true cm
\font\msim=msym10
\def\gr{\hbox{\msim R}}
\def\grt{\hbox{\msim R$^3$}}
\def\grq{\hbox{\msim R$^4$}}
\def\grh{\hbox{\msim R$^8$}}
\def\grc{\hbox{\msim R$^5$}}
\def\grh{\hbox{\msim R$^8$}}
\def\ce{\hbox{\msim C}}

\rightline{\bf LYCEN 9110}
\rightline{(April 1991)}

\sb
\sa

\centerline{\bf ON HURWITZ TRANSFORMATIONS}

\sb
\sa

\centerline{{\bf M. HAGE HASSAN}$^{1}$ and {\bf M. KIBLER}$^2$}

\sa

\centerline{$^1$Universit\'e Libanaise}

\centerline{Facult\'e des Sciences, Section 1}

\centerline{Hadath, Beyrouth, Lebanon}

\sa

\centerline{$^2$Institut de Physique Nucl\'eaire de Lyon}

\centerline{IN2P3~-~CNRS et Universit\'e Claude Bernard}

\centerline{43, Boulevard du 11 Novembre 1918}

\centerline{69622 Villeurbanne Cedex, France}

\sb
\sb
\sb

\noindent
Paper published in the proceedings of the workshop
``Le probl\`eme de factorisation de Hurwitz:
approche historique, solutions, applications en physique'',
Eds., A.~Ronveaux and D.~Lambert (Facult\'es Universitaires
N.D.~de la Paix, Namur, 1991). pp.~1-29.

\vfill\eject

\baselineskip = 0.75 true cm

\centerline{\bf ON HURWITZ TRANSFORMATIONS}

\sb
\sb

\baselineskip = 0.44 true cm

\centerline{{\bf M. HAGE HASSAN}$^{1}$ and {\bf M. KIBLER}$^2$}

\sa

\centerline{$^1$Universit\'e Libanaise}

\centerline{Facult\'e des Sciences, Section 1}

\centerline{Hadath, Beyrouth, Lebanon}

\sa

\centerline{$^2$Institut de Physique Nucl\'eaire de Lyon}

\centerline{IN2P3~-~CNRS et Universit\'e Claude Bernard}

\centerline{43, Boulevard du 11 Novembre 1918}

\centerline{69622 Villeurbanne Cedex, France}

\sb
\sa

\centerline{ABSTRACT}

\vskip 0.28 true cm
\baselineskip = 0.46 true cm

\leftskip = 1.5 true cm
\rightskip = 1.5 true cm

\noindent A bibliography on the Hurwitz transformations is
given. We deal here, with some details, with two particular
Hurwitz transformations, viz, the $\grq \to \grt$
Kustaanheimo-Stiefel transformation and its $\grh \to \grc$
compact extension. These transformations are derived in the
context of Fock-Bargmann-Schwinger calculus with special
emphasis on angular momentum theory.

\leftskip = 0 true cm
\rightskip = 0 true cm

\baselineskip = 0.75 true cm

\sa
\sb
\vskip 0.2 true cm
\vskip 0.2 true cm
\vskip 0.2 true cm

\centerline {\bf 1. Introduction}
\bigskip
The subject of this paper differs from the one (Hurwitz
transformations and quantum groups) of the oral presentation
given by one of us (M.~K.) at the workshop. In fact, although
very interesting, the use of non-bijective canonical
transformations (like the Hurwitz transformations) in
connection with quantum groups is a rather limited subject as
far as the Hurwitz factorization problem is concerned.
Therefore, the present work is devoted to some non-bijective
canonical transformations directly connected to the Hurwitz
factorization problem.
\smallskip
A prototype of non-bijective canonical transformations is
provided by the (now so-called) Kustaanheimo-Stiefel (KS)
transformation [1]. Indeed, this transformation is
inherent to the Hopf fibration $S^3/S^1 = S^2$ [2]
and is closely related to the theory of Cartan
spinors [1,3,4] and, therefore, to the
theory of Hamilton quaternions. Since the basic works of
Refs.~[1,4], the KS transformation and more general non-bijective
canonical transformations have been the
object of numerous studies [5-23]. In particular, the algebra
of (ordinary) quaternions [1,6,7,9,11,12] and,
more generally, the Cayley-Dickson algebras of (ordinary and
hyperbolic) hypercomplex numbers [10,16] proved to be an appropriate
framework for deriving quadratic transformations which extend
the Levi-Civita [1] and the KS transformations. In
addition, it has been shown [22] that the use of Cayley-Dickson algebras
also allows to generate non-quadratic transformations as for example
the Fock [24] (stereographic) transformation.
\smallskip
According to Lambert and Kibler [16], an Hurwitz
transformation is defined as follows. Let ${\cal A}(c)$ be a
Cayley-Dickson algebra of dimension $2m$ and $j$ an
anti-involution of ${\cal A}(c)$. (The abbreviation $(c)$ stands for
$(c_1)$, $(c_1,c_2)$, $(c_1,c_2,c_3)$, $\ldots$
for the algebras of complex numbers, quaternions, octonions,
$\ldots$, respectively, where the $c_i$'s are $\pm1$.) The map
$$
{\cal A}(c) \to {\cal A}(c) \ : \ u \mapsto x = u \, j(u)
\eqno (0a)
$$
defines a (right) Hurwitz transformation which can be described
in matrix form by
$$
{\bf x} = A({\bf u}) \, \varepsilon \, {\bf u} \eqno (0b)
$$
where $\varepsilon$ is a $2m \times 2m$ matrix describing the
anti-involution $j$ and $A({\bf u})$ a matrix
generalizing the ones encountered in the Hurwitz factorization
problem which originally corresponds to the cases
$(2m = 2 \, ; \, c_1 = - 1)$,
$(2m = 4 \, ; \, c_1 = - 1 \, , \, c_2 = - 1)$,
$(2m = 8 \, ; \, c_1 = - 1 \, , \, c_2 = - 1 \, , \, c_3 = - 1)$.
The KS transformation is associated to
$(2m = 4 \, ; \, c_1 = - 1 \, , \, c_2 = - 1)$. It is to be
noticed that, besides the Hurwitz transformations defined by
equation (0), there are other quadratic transformations (the
so-called quasi-Hurwitz and pseudo-Hurwitz transformations [16])
which can be obtained by replacing $\varepsilon$ by an
euclidean or pseudo-euclidean metric matrix. Furthermore,
non-quadratic transformations can be obtained [22] when
replacing $A({\bf u})$ by $A({\bf u})^N$ in
equation (0) ; the geometrical interpretation of the
latter transformations is an interesting problem. The reader
may consult the papers by Lambert and Randriamihamison in these
proceedings for other transformations in an enlarged Hurwitz context.
\smallskip
It is the aim of this work to generate the ($\grq \to \grt$)
KS transformation and its $\grh \to \grc$ compact extension,
corresponding to the Hurwitz transformation with
$(2m = 8 \, ; \, c_1 = -1 \, , \, c_2 = -1 \, , \, c_3 = -1)$,
in the framework of Fock-Bargmann-Schwinger [25-27] calculus.
In this framework, the vector fields
associated to the $\grh \to \grc$ and $\grq \to \grt$ transformations acquire a
nice significance in terms of angular momenta. The material reported here
thus sheds light on the Hurwitz transformations in a new
direction. In particular, it might be useful for the definition
of Hurwitz transformations in the context of quantum matrix
groups \`a la Woronowicz [28] ; as a matter of fact, the
presentation adopted here makes use of a matrix realization of
$SU(2)$ that might be easily deformed in order to switch to
the quantum group $SU_q(2)$.
{}From a physical point of view, the formalism
described in the present paper is clearly
of interest for the connection between the hydrogen atom
(in $\grt$ or $\grc$) and
the isotropic harmonic oscillator (in $\grq$ or $\grh$) as
well as for the separation between vibration and rotation motions.
The reader may consult the paper by Campigotto in these
proceedings for a recent application of the KS transformation
to the hydrogen-oscillator connection.
\smallskip
The paper is organized in the following way. Section 2 is devoted to some
preliminaries. In section 3 and 4, we deal with the compact
Hurwitz transformations
$\grq \to \grt$ and $\grh \to \grc$. Appendices 1 and 2 contain
some elements on Fock-Bargmann-Schwinger calculus.
Finally, some differential aspects of the considered Hurwitz
transformations are relegated to appendices 3 and 4.
\smallskip
The present work constitutes a revision and an extension of the
one contained in a preliminary note published in Ref.~[23].
\smallskip
One of the authors (M.~K.) acknowledges A. Ronveaux and D.
Lambert for inviting him to give a talk at the workshop. He is
also grateful to M.D. Vivarelli and B. Cordani for sending him
articles prior to publication.
\smallskip
\bigskip
\centerline {\bf 2. Preliminaries}
\bigskip
A basic ingredient of this work is the generating function $\Psi$ for the
rotation matrix elements ${\cal D}^j({\hat U})_{mm'}$ of the group
$SU(2)$. To establish the notation, let $\Psi$ be the function
defined by (cf. Bargmann [26])
$$
\Psi(a_1,a_2,b_1,b_2;z_1,z_2) = \exp ({\tilde a}Ub) \qquad
{\tilde a} = (a_1a_2) \qquad
U = \pmatrix{
z_1&z_2\cr
-{\bar z_2}&{\bar z_1}\cr
} \qquad
b = \pmatrix{
b_1\cr
b_2\cr
}
\eqno (1)
$$
The matrix ${\hat U} = r^{-1/2}U$,
with $r = z_1{\bar z_1} + z_2{\bar z_2}$, belongs to
$SU(2)$ and $\Psi$ can be expanded as
$$\Psi(a_1,a_2,b_1,b_2;z_1,z_2) =
\sum_{jmm'} \, \Phi_{jm}(a_1,a_2) \; r^j \; {\cal D}^j({\hat U})_{mm'} \;
\Phi_{jm'}(b_1,b_2)
\eqno (2)$$
where the function $\Phi_{jm}$ is defined in the Fock-Bargmann space
${\cal F}_2$ via
$$\Phi_{jm}(\zeta_1,\zeta_2) =
{{\zeta_1^{j+m}\, \zeta_2^{j-m}}\over
{\sqrt{(j+m)! \, (j-m)!}}}
\eqno (3)$$
In the main body of this paper we shall use the
coordinates $(\sqrt r, \psi, \theta, \varphi)$ of $\grq$ defined through
$$
z_1 = \sqrt r \, \cos {\theta \over 2} \;
\exp (i{{\psi + \varphi} \over 2}) \qquad
z_2 = \sqrt r \, \sin {\theta \over 2} \;
\exp (i{{\psi - \varphi} \over 2})
\eqno (4)$$
i.e., the Euler-angle (quaternionic) coordinates.
\smallskip
The whole philosophy of our approach can be
summarized as follows. Starting from the relation
$$P_q \, q' = 0
\eqno(5)$$
valid for independent coordinates $q$ and $q'$ (with $P_q = {\partial \over
\partial q})$, we look for invariants $\{ q' \}$ with respect to some group
with
infinitesimal operators $\{ P_q \}$. The coordinates $\{q,q'\}$ chosen
are precisely those simple combinations of the quaternionic coordinates
of $\grq$ which are especially adapted to an $SO(4)$ presentation
of the theory of angular momentum.
\smallskip
\bigskip
\centerline {\bf 3. The $\grq \to \grt$ case}
\bigskip
{\bf 3.1. The $\grq \to \grt$ map.} Let
$(\sqrt{r}, \psi, \theta, \varphi)$ be the curvilinear
coordinates of \grq. We choose
$$x_1 = r \sin \theta \cos \varphi \qquad x_2 = r \sin \theta \sin \varphi
\qquad x_3 = r \cos \theta \eqno (6)$$
for the coordinates of \grt. Therefore, our starting point is
$$P_\psi \, x_i = 0 \qquad (i = 1,2,3) \eqno (7)$$
for the choice of coordinates $(x_1,x_2,x_3,\psi)$ in \grq. Since
$P_\psi \, {\cal D}^j({\hat U})_{mm'} = 0$ for $m = 0$,
it is natural to consider the series
$$S_b = \sum_{\ell m'} \, r^\ell \, {\cal D^\ell} ({\hat U})_{0m'} \,
\Phi _{\ell m'}(b_1,b_2) \eqno (8)$$
especially in view of its importance for generating basis functions of the
group $SO(3)$.\par
The expression $S_b$ can be derived by integrating in the Fock-Bargmann space
${\cal F}_2$ the function $\Psi$, in the form given by (2),
multiplied by a convenient factor. As a point of fact, it is immediate to show
that
$$
\int_{\ce} \, \int_{\ce} \, \exp({\bar a_1 \bar a_2}) \,
\Psi(a_1,a_2,b_1,b_2 ; z_1,z_2) \, d\mu(a_1) \, d\mu(a_2) = S_b
\eqno (9)
$$
by using the basic integral of Appendix 1. Now, we pass to
the form (1) for the function $\Psi$ and easily obtain
$$
\int_{\ce} \, \int_{\ce} \, \exp({\bar a_1 \bar a_2}) \,
\Psi(a_1,a_2,b_1,b_2 ; z_1,z_2) \, d\mu(a_1) \, d\mu(a_2) =
\exp[{(z_1b_1+z_2b_2) \, (-\bar z_2b_1 + \bar z_1b_2)}] \eqno (10)$$
after repeated use of the composition rule
of Appendix 1. As an intermediate result, from equations (8-10) we have
$$
\exp [(z_1b_1 + z_2b_2) \, (-\bar z_2b_1 + \bar z_1b_2)]
= \sum _{\ell m'} \; r^\ell \, {\cal D}^\ell ({\hat U})_{0m'} \,
\Phi_{\ell m'} (b_1,b_2) \eqno (11)
$$
On another hand, we know that the generating function for the
$SO(3) \supset SO(2)$ solid harmonics
$$y_{\ell m}(\vec r) =
{\sqrt{{2\ell + 1} \over {4 \pi}}} \, r^\ell \,
{\cal D}^\ell ({\hat U})_{0m} \eqno (12)$$
is given by (cf. Schwinger [27])
$$\exp \left( - {{x_1+ix_2}\over 2} b^2_1 +
                {{x_1-ix_2}\over 2} b^2_2 + x_3b_1b_2\right) =
\sum_{\ell m} \, {\sqrt{{4\pi} \over {2\ell +1}}} \,
y_{\ell m} (\vec r) \, \Phi_{\ell m} (b_1,b_2) \eqno (13)$$
with $\vec r = (x_1,x_2,x_3)$. Introducing (12) into
(13) and identifying the so-obtained relation to (11), we finally get
$$x_1 = z_1 \bar z_2 + \bar z_1 z_2\qquad
x_2 = - i (z_1 \bar z_2 - \bar z_1 z_2)\qquad
x_3 = z_1 \bar z_1 - z_2 \bar z_2 \eqno (14)$$
The latter relations are a simple rewriting of three of the relations
defining the KS transformation. Indeed, by putting
$$z_1 = u_1 + iu_2 \qquad z_2 = u_3 + iu_4 \eqno (15)$$
equation (14) can be rewritten in terms of the Cartesian coordinates
$(u_1,u_2,u_3,u_4)$ of {\grq} as
$$x_1 = 2(u_1u_3 + u_2u_4)\qquad
x_2 = 2(-u_1u_4 + u_2u_3)\qquad
x_3 = u^2_1 + u^2_2 - u^2_3 - u^2_4 \eqno (16)$$
with the evident property that $r = \left(\sum ^3_{i=1} x_i^2 \right)^{1/2} =
\sum ^4_{\alpha = 1} u_\alpha ^2$.
\smallskip
The $\grq \to \grt$ differentiable map defined by (16) corresponds to the
transformation originally introduced by Kustaanheimo and Stiefel [1] up to
permutations on the $x_i$'s and the $u_\alpha$'s. The compatibility between
(4), (6) and (14) gives to (16) its specific form owing to
(15). In the terminology of Lambert and Kibler [16]
and of Kibler and Winternitz [19],
the transformation (16) can be identified to an Hurwitz
transformation of type (c$^{\prime}$)
associated to a certain anti-involution of the Cayley-Dickson algebra
${\cal A}(-1,-1)$, which is isomorphic to the algebra of Hamilton quaternions.
\smallskip
Should we have chosen to work with the series
$$
\eqalign{
S_a & = \sum_{\ell m} \, \Phi_{\ell m}(a_1,a_2) \, r^\ell \,
{\cal D}^\ell {({\hat U})}_{m0}\cr
    & = \int_{\ce} \int_{\ce} \, \exp({\bar b_1 \bar b_2}) \,
\Psi (a_1,a_2,b_1,b_2 ; z_1,z_2) \, d\mu(b_1) \, d\mu (b_2)
} \eqno (17)
$$
instead of $S_b$, we would have arrived at the transformation
$$\eqalign{
x'_1&= z_1z_2 + \bar z_1\bar z_2 = 2(u_1u_3 -u_2u_4)\cr
x'_2&= - i(z_1z_2 - \bar z_1 \bar z_2) = 2 (u_1u_4 + u_2u_3)\cr
x'_3&= z_1 \bar z_1 - z_2 \bar z_2 = u^2_1 + u^2_2 - u^2_3 - u^2_4\cr
} \eqno (18)$$
which corresponds, in the notation of Ref.~[16], to the
Hurwitz transformation associated to the
anti-involution $j_1$ of ${\cal A}(-1,-1)$ through the
relation ${\bf x}' = A({\bf u}) \, \varepsilon_1 \, {\bf u}$,
up to a re-labeling in ${\bf x}'$ and ${\bf u}$.
Equations (4) and (18) can be combined to yield
$$ x'_1 = r \sin \theta \cos \psi \qquad
x'_2 = r \sin \theta \sin \psi \qquad
x'_3 = r \cos \theta \eqno (19)$$
and thus the transformation (18) is inherent
to an approach the starting point of which is
$$
P_\varphi x'_i = 0 \qquad (i = 1,2,3) \eqno (20)
$$
for the choice of coordinates $(x'_1,x'_2,x'_3, \varphi)$ in {\grq}.
\smallskip
Of course, the transformation (18) is equivalent to the transformation
(16). Both transformations are associated to the Hopf fibration on spheres
$S^3/S^1 = S^2$ with compact fiber $S^1$. The reader will find in Appendix
3 some differential aspects of the KS transformation defined by (16).
\smallskip
{\bf 3.2. The {$\grq \to \grt$} vector fields.} The next step
is to examine, still in the framework of Fock-Bargmann-Schwinger
calculus, the implications of (16) and (18) in the language of
angular momentum theory. The basic relations are (see
Refs.~[29,30] and Appendix 2)
$${\hat K} \Psi = K ^{\dag} \Psi \qquad {\hat L}\Psi = L^{\dag} \Psi
\eqno (21)$$
where ${\hat K} = {\hat K}_+$, ${\hat K}_z$ or ${\hat K}_-$
  and ${\hat L} = {\hat L}_+$, ${\hat L}_z$ or ${\hat L}_-$
are the images in ${\cal F}_2$ of $K = K_+$, $K_z$ or $K_-$ and
                                  $L = L_+$, $L_z$ or $L_-$ in the
variables $(a_1,a_2)$ and $(b_1,b_2)$, respectively. The operators
${\vec K} = (K_+, K_z, K_-)$ and
${\vec L} = (L_+, L_z, L_-)$ generate the Lie
algebra of the group $SO(4)$. Let us
simply recall that ${\cal D}^j({\hat U})_{mm'}$ is
an eigenvector of the operators $K_z$ and $L_z$ with the eigenvalues $m$ and
$m'$, respectively. The Fock-Bargmann representations of $\vec K$ and $\vec L$
are given by
$$\eqalign{
{\hat K}_+ &= - a_1 {\partial \over {\partial a_2}}\cr
{\hat L}_+ &= + b_1 {\partial \over {\partial b_2}}\cr
}\quad
\eqalign{
{\hat K}_z &= {1 \over 2} \left (a_1 {\partial \over {\partial a_1}}
- a_2 {\partial \over {\partial a_2}}\right)\cr
{\hat L}_z &= {1 \over 2} \left (b_1 {\partial \over {\partial b_1}}
- b_2 {\partial \over {\partial b_2}}\right)\cr
}\quad
\eqalign{
{\hat K}_- &= - a_2 {\partial \over {\partial a_1}}\cr
{\hat L}_- &= + b_2 {\partial \over {\partial b_1}}\cr
} \eqno (22)$$
and act on the space ${\cal F}_2\otimes{\cal F}_2$.
\smallskip
We begin with the image $\hat K$ of $K$. From straightforward calculations,
we have
$${\hat K}_+ \Psi = - \left(\bar z_1 {\partial \over {\partial z_2}}
- \bar z_2 {\partial \over {\partial z_1}}\right)\Psi
\qquad
  {\hat K}_- \Psi = + \left(z_1 {\partial \over {\partial {\bar z}_2}}
- z_2 {\partial \over {\partial {\bar z}}_1}\right)\Psi$$
$${\hat K}_z \Psi = {1\over 2} \left(z_1 {\partial \over {\partial z_1}}
- {\bar z}_1 {\partial \over \partial {\bar z}_1}
+ z_2 {\partial \over {\partial z_2}}
- \bar z_2 {\partial \over \partial {\bar z}_2}\right) \Psi \eqno (23)$$
Therefore, from (21) we obtain the three spherical components of $\vec K$
$$K_+ = + \left(z_1 {\partial \over {\partial \bar z}_2} -
z_2 {\partial \over {\partial {\bar z}} _1}\right)
\qquad
K_- =  - \left({\bar z}_1 {\partial \over {\partial z_2}} -
{\bar z}_2 {\partial \over {\partial z_1}}\right)$$
$$K_z = {1 \over 2} \left(z_1 {\partial \over \partial z_1} -
{\bar z}_1 {\partial \over {\partial {\bar z}_1}} +
z_2 {\partial \over {\partial z_2}} -
{\bar z}_2 {\partial \over {\partial {\bar z}_2}}\right) \eqno (24)$$
With the help of (15) and of the well-known relations
$${\partial \over {\partial z_1}} =
{1 \over 2} \left(\partial_1 - i \partial_2\right)\qquad
  {\partial \over {\partial z_2}} =
{1 \over 2} \left(\partial_3 - i \partial_4\right) \eqno (25)$$
where $\partial_\alpha$ stands for $\partial \over {\partial u_\alpha}$, we can
derive from (24) the expressions of $K_+$, $K_z$ and $K_-$ in terms of the
coordinates $u_\alpha$ $(\alpha = 1,2,3,4)$. Finally, by introducing the
Cartesian components
$$K_1 = {1 \over 2} (K_+ + K_-) \qquad
K_2 = {1 \over {2i}} (K_+ - K_-) \qquad
K_3 = K_z \eqno (26)$$
we get the vector fields
$$\eqalign {
K_1 &= - {i \over 2} (+ u_4 \partial_1 + u_3 \partial_2
- u_2 \partial_3 - u_1 \partial_4)\cr
K_2 &= - {i \over 2} (- u_3 \partial_1 + u_4 \partial_2
+ u_1 \partial_3 - u_2 \partial_4)\cr
K_3 &= + {i \over 2} (+ u_2 \partial_1 - u_1 \partial_2
+ u_4 \partial_3 - u_3 \partial_4)} \eqno (27)$$
defined in the real symplectic Lie algebra $sp(8,{\gr})$.
\smallskip
In a similar fashion, starting from ${\hat L} \Psi = L^{\dag} \Psi$,
we would obtain the following spherical components of ${\vec L}$
$$
  L_+ = + \left(z_1 {\partial \over {\partial z_2}} -
{\bar z}_2 {\partial \over {\partial {\bar z}_1}}\right) \qquad
  L_- = - \left({\bar z}_1 {\partial \over {\partial \bar z_2}}
- z_2 {\partial \over {\partial z_1}}\right)
$$
$$
  L_z = {1 \over 2} \left(z_1 {\partial \over \partial z_1}
- {\bar z}_1 {\partial \over \partial {\bar z}_1}
- z_2  {\partial \over \partial z_2}
+ {\bar z}_2 {\partial \over \partial {\bar z}_2}\right)
\eqno (28)
$$
so that
$$\eqalign {
L_1 &= {i \over 2} \left(u_4 \partial_1 - u_3 \partial_2
+ u_2 \partial_3 - u_1 \partial _4\right)\cr
L_2 &= {i \over 2} \left(u_3 \partial_1 + u_4 \partial_2
-u_1 \partial_3 - u_2 \partial_4\right)\cr
L_3 &= {i \over 2} \left(u_2 \partial_1 - u_1 \partial_2
- u_4 \partial_3 + u_3 \partial_4\right)\cr
} \eqno (29)$$
are the three Cartesian components of $\vec L$ expressed in
the algebra $sp(8,{\gr})$.
\smallskip
It is easy to verify that the commutation relations
$$\left[K_k, K_{\ell }\right] = i \, \epsilon _{k\ell m} \, K_m
\qquad
  \left[L_k, L_{\ell }\right] = i \, \epsilon _{k\ell m} \, L_m
\qquad
\left[K_k, L_\ell \right] = 0 \eqno (30)$$
hold for $k, {\ell}, m = 1$, $2$ or $3$. Consequently, the set $\{K_j,L_j : j =
1,2,3 \}$ spans the Lie algebra $so(4)$ in an $su(2) \oplus su(2)$ basis, as
expected from the angular momentum theory developed in an $SO(4)$ presentation
(cf. Ref.~[31]).
\smallskip
At this stage, it should be noted that the vector operators $\vec K$ and
$\vec L$ are angular momenta associated to the coordinates $(x'_1,x'_2,x'_3)$
and $(x_1,x_2,x_3)$, respectively. More precisely, it is a simple matter of
calculation to show that the components of $\vec K$ can be written
$$K_1 = + i \left(x'_2 {\partial \over {\partial x'_3}}
- x'_3 {\partial \over {\partial x'_2}}\right)
= - i \left(+ \sin \psi {\partial \over {\partial \theta}}
+ \cot \theta \cos \psi {\partial \over {\partial \psi}}\right)$$
$$K_2 = + i \left(x'_3 {\partial \over {\partial x'_1}}
- x'_1 {\partial \over {\partial x'_3}}\right)
= - i \left(- \cos \psi {\partial \over {\partial \theta}}
+ \cot \theta \sin \psi {\partial \over {\partial \psi}}\right)$$
$$K_3 = - i \left({x'_1 {\partial \over {\partial x'_2}}
- x'_2 {\partial \over {\partial x'_1}}}\right)
= - i {\partial \over {\partial \psi}} \eqno (31)$$
Similarly, the components of $\vec L$ satisfy
$$ L_1 = - i \left({x_2 {\partial \over {\partial x_3}}
- x_3 {\partial \over {\partial x_2}}}\right)
= i \left({+ \sin \varphi {\partial \over {\partial \theta}}
+ \cot \theta \cos \varphi {\partial \over {\partial \varphi}}}\right)$$
$$L_2 = - i \left({x_3 {\partial \over {\partial x_1}}
- x_1 {\partial \over {\partial x_3}}}\right)
= i \left({- \cos \varphi {\partial \over {\partial \theta}}
+ \cot \theta \sin \varphi {\partial \over {\partial \varphi}}}\right)$$
$$L_3 = - i \left({x_1 {\partial \over {\partial x_2}}
- x_2 {\partial \over {\partial x_1}}}\right)
= - i {\partial \over {\partial \varphi}} \eqno (32)$$
and are nothing but the components of the ordinary angular
momentum in the coordinates $(x_1,x_2,x_3)$. Further,
we note that in order to pass from the operator $\vec K$ defined by
(24) to the operator $\vec L$ defined by (28) it is sufficient to replace
$z_2$ by ${\bar z}_2$. This reflects the fact that we can pass from the
coordinates $(x'_1, x'_2, x'_3)$ to the coordinates $(x_1, x_2, x_3)$ by
changing $z_2$ into ${\bar z}_2$.
\smallskip
Finally, it should be observed that the constraint operator $X$
discussed by many authors (see, for example, Refs.~[5,8,16]) is
$$
X = {2 \over i} \, K_3 = - 2 P_{\psi} =
u_2 \partial_1 - u_1 \partial_2 + u_4 \partial_3 - u_3 \partial_4
\eqno (33)
$$
for the transformation (16) and
$$
X = {2 \over i} \, L_3 = - 2 P_{\varphi} =
u_2 \partial_1 - u_1 \partial_2 - u_4 \partial_3 + u_3 \partial_4
\eqno (34)
$$
for the transformation (18). In this respect, the constraint relation
$$X f = 0 \eqno (35)$$
holds either
for $X = (2/i) K_3$ and $f = f(x_1,x_2,x_3)$ or for $X = (2/i)L_3$ and
$f = f(x'_1,x'_2,x'_3)$, where $f$ is a (one-fold) differentiable function.
(Equation (35) was the starting point of our derivation of the
transformations (16) and (18).) The operator $X$ generates an $so(2)$
subalgebra, with $so(2) = so(2)_K$ for $X = (2/i)K_3$ and $so(2) = so(2)_L$ for
$X = (2/i)L_3$, of the symplectic Lie algebra $sp(8,{\gr})$. The corresponding
Lie algebra under constraints is
$$
{\rm cent}_{sp(8,{\gr})}(so(2))/so(2) = so(4,2)\sim su(2,2)
\eqno (36)
$$
(see Kibler and N\'egadi [8] and Kibler
and Winternitz [19]). To close this section, note that the
one-form $\omega$ (see Ref.~[16]) associated to the vector field $X$ can be
recovered from the property $\omega \left[{1 \over {2r}} X\right] = 1$.
\smallskip
\bigskip
\centerline {\bf 4. The $\grh \to \grc$ case}
\bigskip
We are now in a position to handle the $\grh \to \grc$ case. To deal
with \grh, we start from two copies of $\grq$ and choose the coordinates
in the following manner.
Let us consider two particles, say, $1$ and $2$. Let $(\psi, \theta,
\varphi)$ be, in this section, the angular coordinates of the collective motion
and $(x_1,x_2,\ldots,x_5)$ the five remaining coordinates
necessary to completely
describe, in terms of the $\grq \oplus \grq$ space, the
position of the two particles. In this case, equation (5) reads
$$
P_\psi x_i = P_\theta x_i = P_\varphi x_i = 0 \qquad
(i = 1,2,\ldots,5) \eqno (37)
$$
Equation (37) may be taken in the form
$$
{\cal K}_1 x_i = {\cal K}_2 x_i = {\cal K}_3 x_i = 0 \qquad
(i = 1,2,\ldots,5) \eqno (38)
$$
where ${\cal K}_1$, ${\cal K}_2$ and ${\cal K}_3$ denote the (Cartesian)
coordinates of the total vector operator ${\cal \vec K} = {\vec K}(1) + {\vec
K}(2)$ for the system of the two particles. (The operators ${\vec K}(1)$ and
                                                           ${\vec K}(2)$
correspond to the operator ${\vec K}$ of section 3 for the particles 1 and 2,
respectively.) We call $SO(3)_{\cal K}$ the Lie group whose infinitesimal
generators are ${\cal K}_1$, ${\cal K}_2$ and ${\cal K}_3$.
\smallskip
The function
$$\Psi_{uc} = \Psi_1 \Psi_2 \eqno (39)$$
where $\Psi _1$ and $\Psi _2$ stand for generating functions of type $\Psi$
(see section 3) associated to the particles $1$ and $2$,
respectively, is simply the
uncoupled generating function corresponding to the system of the two particles.
More precisely, we take
$$
\eqalign{
\Psi_1 &= \exp ({\tilde a} Ub) =
\sum _{j_1m_1m'_1} \, \Phi_{j_1 m_1} (a_1,a_2) \,
r_1^{j_1} \, {\cal D}^{j_1} ({\hat U})_{m_1m'_1} \, \Phi_{j_1m'_1} (b_1,b_2)\cr
\Psi_2 &= \exp ({\tilde c} Vd) =
\sum _{j_2m_2m'_2} \, \Phi_{j_2 m_2} (c_1,c_2) \,
r_2^{j_2} \, {\cal D}^{j_2} ({\hat V})_{m_2m'_2} \, \Phi_{j_2m'_2} (d_1,d_2)
} \eqno (40)
$$
where ${\tilde a}$, $U$ and $b$ are defined via equation (1) and similarly
$${\tilde c} = (c_1c_2) \qquad
V = \left(
\matrix{
z_3&z_4\cr
-{\bar z}_4&{\bar z}_3\cr
}
\right) \qquad
d = {d_1\choose d_2} \eqno (41)$$
Furthermore, the rotational invariants $r_1$ and $r_2$ occurring in
(40) are given by
$$r_1 = z_1{\bar z}_1 + z_2{\bar z}_2 \qquad
r_2 = z_3{\bar z}_3 + z_4{\bar z}_4 \eqno (42)$$
Following Bargmann [26] and Schwinger [27], the coupled generating
function $\Psi_c$ for the two-particle system can be taken as
$$
\Psi_c = \int_{\ce^2} \int_{\ce^2} \;
\exp[{\alpha({\bar c}_1 {\bar \eta} - {\bar c}_2 {\bar \xi})
+ \beta ({\bar \xi} {\bar a}_2 - {\bar \eta}{\bar a}_1)
+ \gamma ({\bar a}_1 {\bar c}_2 - {\bar a}_2 {\bar c}_1)}] \;
\Psi_{uc} \; d\mu_2(a) \; d\mu_2(c) \eqno (43)
$$
Now, let us introduce the generating function (see Ref.~[27])
$$
e^{\alpha (b_1 c_2 - b_2 c_1) + \beta (c_1 a_2 - c_2 a_1)
 + \gamma (a_1 b_2 - a_2 b_1)}
= \sum _{j_1 j_2 j_3} \, \sum _{m_1 m_2 m_3}
N_{j_1j_2j_3}
\pmatrix{
j_1&j_2&j_3\cr
m_1&m_2&m_3\cr
}
$$
$$\alpha^{-j_1+j_2+j_3}\;
   \beta^{j_1-j_2+j_3}\;
  \gamma^{j_1+j_2-j_3}\;
\Phi_{j_1m_1} (a_1,a_2)\;
\Phi_{j_2m_2} (b_1,b_2)\;
\Phi_{j_3m_3} (c_1,c_2)
$$
$$
N_{j_1j_2j_3} =
\left[{(j_1+j_2+j_3+j+1)!}
\over
{(-j_1+j_2+j_3)! \; (j_1-j_2+j_3)! \; (j_1+j_2-j_3)!}
\right]^{1 \over 2} \eqno (44)
$$
for the $3-jm$ symbols of the group $SU(2)$ in an $SU(2) \supset U(1)$ basis.
Then, the exponential in (43) can be transformed owing to (44) and the
integration performed by expanding $\Psi_{uc}$ in terms of
$\Phi_{jm}$'s and by
using the orthonormality property of the functions $\Phi_{jm}$ (see
Appendix~1).
We thus obtain
$$\Psi_c =
\sum_{j_1j_2j_3}\,
\sum_{m_1m_2m_3}\,
\sum_{m'_1m'_2}\,
N_{j_1j_2j_3}\,
\alpha^{-j_1+j_2+j_3}\,
\beta^{j_1-j_2+j_3}\,
\gamma^{j_1+j_2-j_3}
\pmatrix{
j_1&j_2&j_3\cr
m_1&m_2&m_3\cr
}$$
$$r_1^{j_1} \, {\cal D}^{j_1} ({\hat U})_{m_1m'_1} \,
  r_2^{j_2} \, {\cal D}^{j_2} ({\hat V})_{m_2m'_2} \,
          \Phi_{j_1m'_1} (b_1,b_2) \,
          \Phi_{j_2m'_2} (d_1,d_2) \,
\overline{\Phi_{j_3 m_3} (\xi,\eta)} \eqno (45)$$
The $SO(3)_{\cal K}$ rotational invariance depicted by (38)
is ensured if we assume that $\alpha = \beta = 0$ in (45). Then,
by taking $\alpha = \beta = 0$
and by introducing (39) and (40) into (43), the integration can be easily
performed after repeated use of the composition rule of Appendix 1. This leads
to
$$\Psi_c = \exp \left \{
\gamma [(z_1b_1+z_2b_2) (-{\bar z}_4d_1 + {\bar z}_3d_2)
- (- {\bar z}_2b_1 + {\bar z}_1b_2) (z_3d_1 + z_4d_2)]
\right \} \eqno (46)$$
On the other hand, we may introduce $\alpha = \beta = 0$ in (45). This yields
$$\Psi_c =
\sum_{JMM'} \, \gamma^{2J} \, \Phi_{JM} (-b_2,b_1) \, (r_1r_2)^J
\, {\cal D}^J ({\hat U}^{\dag} {\hat V})_{MM'} \, \Phi_{JM'}(d_1,d_2)
\eqno (47)$$
which can be obtained equally well from (43) with $\alpha = \beta = 0$ and
$\Psi_{uc}$ expanded in terms of $\Phi_{jm}$'s. Without loss of generality, we
can assume that $\gamma = 1$. Therefore, equations (46) and
(47) show that $\Psi_c$ is of the form (cf. (1) and (41))
$$\Psi_c = \exp (\tilde b' Wd) \qquad
{\tilde b'} = (-b_2 b_1) \qquad
W = \left(
\matrix{
+s_1 + is_2&s_3 + is_4\cr
-s_3 + is_4&s_1 - is_2\cr
}
\right) \qquad
d = {d_1\choose d_2}
\eqno (48)
$$
and we have to identify W with $U^{\dag} V$. As a result, we obtain the four
non-trivial rotational invariants $x_i = 2s_i$ $(i=1,2,3,4)$ given by
$${1 \over 2} (x_1 + ix_2) = {\bar z}_1 z_3 + z_2 {\bar z}_4 \qquad
{1 \over 2} (x_3 + ix_4) = {\bar z}_1 z_4 - z_2 {\bar z}_3
\eqno (49)$$
(The particular normalization for $s_i = (1/2) x_i$ is chosen
in view of having the
property $r = (\sum_{i=1}^5 \, x^2_i)^{1/2} = \sum_{\alpha = 1}^8
\, u^2_\alpha$, see equation (53) below.)
The fifth rotational invariant $x_5$ is obtained by
looking for an expression ``orthogonal'' to the trivial invariant
$$r = r_1 + r_2 =
z_1 {\bar z}_1 + z_2 {\bar z}_2 + z_3 {\bar z}_3 + z_4 {\bar z_4}
\eqno (50)$$
Thereby, we take
$$
x_5 = r_1 - r_2 = z_1 {\bar z_1} + z_2 {\bar z}_2
                - z_3 {\bar z}_3 - z_4 {\bar z}_4 \eqno (51)
$$
Finally, from (49) we are left with
$$
\eqalign{
x_1 & = z_1 {\bar z}_3 + {\bar z}_1 z_3 + z_2 {\bar z}_4 + {\bar z}_2 z_4
\cr
x_2 & = i (z_1 {\bar z}_3 - {\bar z}_1 z_3 - z_2 {\bar z}_4 + {\bar z}_2 z_4)
\cr
x_3 & = z_1 {\bar z}_4 + {\bar z}_1 z_4 - z_2 {\bar z}_3 - {\bar z}_2 z_3
\cr
x_4 & = i(z_1 {\bar z}_4 - {\bar z}_1 z_4 + z_2 {\bar z}_3 - {\bar z}_2 z_3)
} \eqno (52)
$$
Note that the invariant $r$ depends on $x_1$, $x_2$, $x_3$, $x_4$ and $x_5$
since
$$r^2 = x^2_1 + x^2_2 + x^2_3 + x^2_4 + x^2_5 \eqno (53)$$
The use of the relation (15) for the particle 1 and of the parent relation
$$z_3 = u_5 + iu_6 \qquad z_4 = u_7 + iu_8 \eqno (54)$$
for the particle 2 allows us to rewrite (51) and (52) as
$$
\eqalign{
x_1 & = 2(u_1 u_5 + u_2 u_6 + u_3 u_7 + u_4 u_8) \cr
x_2 & = 2(u_1 u_6 - u_2 u_5 - u_3 u_8 + u_4 u_7) \cr
x_3 & = 2(u_1 u_7 + u_2 u_8 - u_3 u_5 - u_4 u_6) \cr
x_4 & = 2(u_1 u_8 - u_2 u_7 + u_3 u_6 - u_4 u_5) \cr
x_5 & = u^2_1 + u^2_2 + u^2_3 + u^2_4 - u^2_5 - u^2_6 - u^2_7 - u^2_8
} \eqno (55)
$$
that defines an $\grh \to \grc$ map. By putting
$${\vec x} = (x_2, x_3, x_4) \qquad
q_{10} = u_1 \qquad
{\vec q}_1 = (u_2, u_3, u_4) \qquad
q_{20} = u_5 \qquad
{\vec q}_2 = (u_6, u_7, u_8) \eqno (56)$$
the relations (55) can be recast in the compact form (see also
Cordani [21])
$$
x_1 = 2 (q_{10} q_{20} + {\vec q}_1 . \, {\vec q}_2) \qquad
{\vec x} = 2(q_{10} {\vec q}_2
- q_{20} {\vec q}_1 - {\vec q}_1 \wedge {\vec q}_2) \qquad
x_5 = q_{10}^2 + {\vec q}_1 \, ^2 - q_{20}^2 - {\vec q}_2 \, ^2
\eqno (57)
$$
which obviously reflects
the fact that our approach is connected with bi-quaternions.
\smallskip
According to Lambert and Kibler [16], the latter map is associated to a
certain Hurwitz transformation. The three corresponding constraint
operators (generalizing the operator $X$ of the KS transformation) are
$$
X_j = {2 \over i} \, {\cal K}_j = {2 \over i} \, [K_j (1) +  K_j (2)]
\qquad (j = 1,2,3) \eqno (58)
$$
and thus acquire a significance in terms of generalized angular momentum. In
this connection, the Lie algebra of $SO(3)\cal_K$ identifies to what Kibler
and N\'egadi [8] and Kibler and Winternitz [19]
call a constraint Lie algebra. Finally, the one-forms
$$
\eqalign{
\omega_1 & = - 2(+ u_4du_1 + u_3du_2 - u_2du_3 - u_1du_4
                 + u_8du_5 + u_7du_6 - u_6du_7 - u_5du_8)\cr
\omega_2 & = - 2(- u_3du_1 + u_4du_2 + u_1du_3 - u_2du_4
                 - u_7du_5 + u_8du_6 + u_5du_7 - u_6du_8)\cr
\omega_3 & = + 2(+ u_2du_1 - u_1du_2 + u_4du_3 - u_3du_4
                 + u_6du_5 - u_5du_6 + u_8du_7 - u_7du_8)
} \eqno(59)
$$
are the duals of $X_1, X_2, X_3$ in the sense that
$\omega_\alpha [{1 \over {2r}} X_\beta] = \delta (\alpha, \beta)$
(see Ref.~[16]).
\smallskip
Similarly, instead of starting from (38),
we could have chosen to look for the
coordinates $x_i'$ $(i = 1, 2, \ldots, 5)$
invariant under the group $SO(3)_{\cal L}$
corresponding to the total angular momentum
${\cal \vec L} = {\vec L} (1) + {\vec
L} (2)$, where ${\vec L} (1)$ and ${\vec L} (2)$
refer to the angular momenta for
the particles 1 and 2, respectively. The basic requirement
would have been
$$
{\cal L}_1 x_i' = {\cal L}_2 x_i' = {\cal L}_3 x_i' = 0 \qquad
(i = 1,2,\ldots,5)
\eqno (60)
$$
a set of constraints equivalent to
$$
P_\varphi x_i' = P_\theta x_i' = P_\psi x_i' = 0 \qquad
(i = 1,2,\ldots,5)
\eqno (61)
$$
Thus, we would have been left with $W' = UV^{\dag}$, i.e.
$$
\eqalign{
x'_1 & = {\bar z}_1 z_3 + z_1{\bar z}_3 + z_2{\bar z}_4 + {\bar z}_2 z_4\cr
x'_2 & = i ({\bar z}_1 z_3 - z_1{\bar z}_3 - z_2{\bar z}_4 + {\bar z}_2 z_4)\cr
x'_3 & = - {\bar z}_1 {\bar z}_4 - z_1z_4 + z_2z_3 + {\bar z}_2 {\bar z}_3\cr
x'_4 & = i (- {\bar z}_1 {\bar z}_4 + z_1 z_4 - z_2 z_3
+ {\bar z}_2 {\bar z}_3)\cr
x'_5 & = z_1 {\bar z_1} + z_2 {\bar z}_2 - z_3 {\bar z}_3 - z_4 {\bar z}_4
} \eqno (62)
$$
and, therefore, with the $\grh \to \grc$ map defined by
$$
\eqalign{
x'_1 & = 2 (+ u_1 u_5 + u_2 u_6 + u_3 u_7 + u_4 u_8) \cr
x'_2 & = 2 (- u_1 u_6 + u_2 u_5 - u_3 u_8 + u_4 u_7) \cr
x'_3 & = 2 (- u_1 u_7 + u_2 u_8 + u_3 u_5 - u_4 u_6) \cr
x'_4 & = 2 (- u_1 u_8 - u_2 u_7 + u_3 u_6 + u_4 u_5) \cr
x'_5 & = u^2_1 + u^2_2 + u^2_3 + u^2_4 - u^2_5 - u^2_6 - u^2_7 - u^2_8
} \eqno (63)
$$
or alternatively (in a quaternionic form) by
$$
x'_1 = 2 (q_{10} q_{20} + {\vec q}_1 .\, {\vec q}_2) \qquad
{\vec {x'}} = 2 (-q_{10} {\vec q}_2 + q_{20} {\vec q}_1
- {\vec q}_1 \wedge {\vec q}_2) \qquad
x'_5 = q^2_{10} + {\vec q}_1 \, ^2 - q_{20}^2 - {\vec q}_2 \, ^2
\eqno (64)
$$
with ${\vec {x'}} = (x'_2, x'_3, x'_4)$. Such a map corresponds,
up to a change of notations, to the Hurwitz
transformation associated to the anti-involution $j_1$ of the algebra of
octonions ${\cal A}(-1,-1,-1)$ through
the relation ${\bf x}' = A({\bf u}) \, \varepsilon_1 \, {\bf u}$
(see Ref.~[16]). Here, the three constraint operators are
$$
X_j = {2 \over i} \, {\cal L}_j = {2 \over i} \, [L_j(1) + L_j(2)]
\qquad (j = 1,2,3) \eqno (65)
$$
and the corresponding one-forms $\omega_j$ ($j = 1, 2, 3$) are
$$
\eqalign{
\omega_1 & = 2(u_4du_1 - u_3du_2 + u_2du_3 - u_1du_4 +
               u_8du_5 - u_7du_6 + u_6du_7 - u_5du_8) \cr
\omega_2 & = 2(u_3du_1 + u_4du_2 - u_1du_3 - u_2du_4 +
               u_7du_5 + u_8du_6 - u_5du_7 - u_6du_8) \cr
\omega_3 & = 2(u_2du_1 - u_1du_2 - u_4du_3 + u_3du_4 +
               u_6du_5 - u_5du_6 - u_8du_7 + u_7du_8)
} \eqno (66)
$$
with the property that $\omega_\alpha \left[{1 \over {2r}} X_\beta\right] =
\delta(\alpha, \beta)$.
\smallskip
The $\grh \to \grc$ maps defined via equations (55) and
(58), on one hand, and via equations (63) and (65), on the
other hand, are identical up to a re-labeling of the various
variables. For both maps, we have
$$X f = 0 \eqno (67)$$
with either $X \in so(3)_{\cal K}$ and $f = f(x_1,x_2, \ldots,x_5)$ or
            $X \in so(3)_{\cal L}$ and $f = f(x'_1,x'_2,\ldots,x'_5)$,
where $f$ is a (one-fold) differentiable function. The constraint
Lie algebra $so(3) = so(3)_{\cal K}$ or $so(3)_{\cal L}$ is
clearly a subalgebra of the real symplectic Lie algebra $sp(16, \gr)$ and the
corresponding Lie algebra under constraints is
$$
{\rm cent}_{sp(16,\gr)}(so(3)) = so(6,2) \sim so^\ast(8) \eqno (68)
$$
(see Ref.~[19]). Indeed, the two maps
under consideration correspond to the Hopf fibration on spheres $S^7/S^3 \to
S^4$ with compact fiber $S^3$. Some further differential aspects of these $\grh
\to \grc$ maps are discussed in Appendix 4.

\vfill\eject

\centerline {\bf Appendix 1: Fock-Bargmann-Schwinger calculus}
\smallskip
\bigskip
Let ${\cal F}_n$ be the Bargmann space of entire analytical
functions $f(\zeta)$, the $n$-uple $\zeta=(\zeta_1,\zeta_2,...,\zeta_n)$
being a point of the $n$-dimensional complex (Euclidean)
space {\msim C}$^n$. Such a space can be endowed with the scalar product
defined
via
$$
(f,f') = \int_{\hbox{\msim C}^n} \,
\overline{f(\zeta)} \; f'(\zeta) \; d\mu_n(\zeta)
$$
where the bar indicates complex conjugation and the measure of integration over
{\msim C}$^n$ is given by
$$
d\mu_n(\zeta) = \prod_{k=1}^n d\mu(\zeta_k) \qquad \quad
                              d\mu(\zeta_k) = {1\over\pi} \;
\exp {(-\bar \zeta_k \zeta_k)} \; dx_k \; dy_k \qquad \quad
\zeta_k = x_k + iy_k
$$
As a particular $(f,f')$, we have
$$
\left( \zeta^{\langle h \rangle} , \zeta^{\langle h' \rangle} \right) =
\delta (\langle h' \rangle , \langle h \rangle) \; \langle h \rangle !
$$
for $f (\zeta) = \zeta^{\langle h  \rangle}$ and
    $f'(\zeta) = \zeta^{\langle h' \rangle}$, where we use the notation
$$
\zeta^{\langle h \rangle} = \zeta_1^{h_1} \zeta_2^{h_2} \ldots \zeta_n^{h_n}
\qquad \quad
\zeta^{\langle h'\rangle} = \zeta_1^{h_1'}\zeta_2^{h_2'}\ldots \zeta_n^{h_n'}
$$
$$
\delta (\langle h' \rangle , \langle h \rangle) =
\delta (h'_1,h_1) \; \delta(h'_2,h_2) \; \ldots \; \delta(h'_n,h_n)
$$
$$
\langle h \rangle ! = h_1! \; h_2! \; \ldots \; h_n ! \qquad \quad
h_k \in \hbox{\msim N} \qquad \quad k = 1,2,\ldots,n
$$
Repeated use of the latter scalar product leads to the composition rule
$$
(e^{\bar\alpha \zeta_k}, e^{\beta \zeta_\ell}) = 1 \quad
\hbox {if } \quad k \ne \ell \quad \hbox {or } \quad  e^{\alpha\beta}
\quad \hbox {if } \quad k = \ell
$$
for any $\alpha$ and $\beta$ in {\msim C}.
\smallskip
{\bf The case $n = 1$.} In this trivial case, we obtain the basic integral
$$
\int _{\hbox {\msim C}} \bar \zeta^h \, \zeta^{h'} \, d\mu(\zeta) =
{1 \over \pi} \int ^{+\infty}_{-\infty} \int ^{+\infty}_{-\infty} \;
(x-iy)^h \; (x+iy)^{h'} \; e^{-(x^2+y^2)} \; dx \; dy = \delta (h',h) \; h!
$$
for the
scalar product $(\zeta^h,\zeta^{h'})$ on ${\cal F}_1$, so that the set
$$
\left \{ \varphi _h(\zeta) = {\zeta^h \over \sqrt{h!}} =
{(x+iy)^h\over\sqrt{h!}} \ : \ h \in \hbox{\msim N} \right \}
$$
constitutes an orthonormal basis for ${\cal F}_1$. Furthermore,
the composition rule
$$
\int _{\hbox{\msim C}} \, e^{\alpha \bar \zeta}
                       \, e^{\beta       \zeta} \, d\mu(\zeta) =
{1\over\pi} \, \int^{+\infty} _{-\infty}
               \int^{+\infty} _{-\infty} \,
e^{\alpha (x-iy)} \, e^{\beta (x+iy)} \, e^{-(x^2+y^2)} \, dx \, dy =
e^{\alpha\beta}
$$
gives the scalar product $(e^{\bar\alpha \zeta},e^{\beta \zeta})$
on ${\cal F}_1$ for any $\alpha$ and $\beta$ in {\msim C}.
\smallskip
{\bf The case $n = 2$.} This case is of special interest for the theory of
angular momentum. Let us consider the function
$$
\Phi_{JM}(\zeta_1,\zeta_2) =
{\zeta_1^{J+M} \; \zeta_2^{J-M}\over \sqrt{(J+M)! \, (J-M)!}}
\qquad \quad 2J \in \hbox{\msim N} \qquad \quad M = - J, - J+1, \ldots, J
$$
It is obvious that
$$
(\Phi_{JM},\Phi_{J'M'}) = \int_{\hbox{\msim C}^2} \,
\overline {\Phi_{JM}(\zeta_1,\zeta_2)} \; \Phi_{J'M'} (\zeta_1,\zeta_2)
\; d\mu_2(\zeta) = \delta(J',J) \; \delta(M',M)
$$
and consequently the set
$$
\left\{\Phi_{JM} \ : \ 2J \in \hbox{\msim N}, \ M = -J, -J+1, \ldots, J\right\}
$$
spans an orthonormal basis for ${\cal F}_2$. It is easy to
find a realization $(\hat J_\pm$ and $\hat J_z)$ in ${\cal F}_2$
of the angular
momentum operators $J_\pm = J_x \pm iJ_y$ and $J_z$. It is a matter of simple
calculation to verify that the operators (with $\hbar = 1$)
$$
\hat J_+ = \zeta_1 {\partial \over \partial \zeta_2} \qquad \quad
\hat J_z = {1\over 2} \left( \zeta_1 {\partial \over \partial \zeta_1}
                           - \zeta_2 {\partial \over \partial \zeta_2}
\right) \qquad \quad
\hat J_- = \zeta_2 {\partial \over \partial \zeta_1}
$$
satisfy the relations
$$\hat J_\pm \Phi_{JM} = \sqrt{J(J+1)-M(M\pm 1)} \, \Phi_{JM\pm1}
\qquad \quad \hat J_z \, \Phi_{JM} = M \, \Phi_{JM}
$$
Operators of the type $\hat J_\pm$ and $\hat J_z$ are employed in
section 3. First,
for $(\zeta_1,\zeta_2)$ = $(b_1,b_2)$, we take $\hat J_\pm = \hat L_\pm$ and
$\hat J_z = \hat L_z$. Second, for $(\zeta_1,\zeta_2) = (a_1,a_2)$, we take
$\hat J_\pm = - \hat K_\pm$ and $\hat J_z = \hat K_z$. (Referring to the change
of sign when going from $L_\pm$ to $K_\pm$, we note that the set
$\{ -J_+, J_z, -J_- \}$ spans the same, viz,
with the same structure constants, $su(2)$ Lie
algebra as the set $\{J_+, J_z, J_-\}$.)
\smallskip
The miscellaneous formulas reported in this appendix are at
the root of Fock-Bargmann-Schwinger calculus [25-27] and are used
for $n = 2$ (with $\zeta = a$ or $b$) and
    $n = 4$ (with $\zeta = a$ and $c$ or $b$ and $d$)
in sections 3 and 4 of the present paper.

\vfill\eject

\centerline {\bf Appendix 2: Images of $\vec L$ and $\vec K$ in ${\cal F}_2$}
\smallskip
\bigskip
This appendix deals with the derivations of equations (21). The basic
relations for the operators $\vec L$ and $\vec K$ are (in units $\hbar = 1$)
$$
L_\pm \, {\cal D}^j(\hat U)_{mm'} = + \sqrt {j(j+1) - m'(m'\pm 1)} \;
{\cal D}^j{(\hat U)}_{m,m'\pm1} \quad \quad L_z \, {\cal D}^j(\hat U)_{mm'} =
m' \, {\cal D}^j (\hat U)_{mm'}
$$
$$
K_\pm \, {\cal D}^j(\hat U)_{mm'} = - \sqrt {j(j+1) - m (m \pm 1)} \;
{\cal D}^j{(\hat U)}_{m\pm1,m'} \quad \quad K_z \, {\cal D}^j(\hat U)_{mm'} =
m  \, {\cal D}^j (\hat U)_{mm'}
$$
We begin with the proof of $\hat L_- \Psi = L_-^\dagger \Psi.$ We have
$$\eqalign{
L_+\Psi & = \sum_{jmm'} \Phi_{jm} (a_1,a_2) \; r^j \;
L_+ {\cal D}^j(\hat U)_{mm'} \; \Phi_{jm'} (b_1,b_2)\cr
& = \sum_{jmm'} \Phi_{jm}(a_1,a_2) \; r^j \; \sqrt{j(j+1) - m'(m'+1)} \;
{\cal D}^j(\hat U)_{m,m'+1} \; \Phi_{jm'}(b_1,b_2)\cr
}$$
On the other hand, from Appendix 1 we get
$$
\eqalign{
\hat L_- \Psi & = \sum_{jmm'} \Phi_{jm}(a_1,a_2) \; r^j \;
{\cal D}^j(\hat U)_{mm'} \; \hat L_- \, \Phi_{jm'}(b_1,b_2)\cr
              & = \sum_{jmm'} \Phi_{jm}(a_1,a_2) \; r^j \;
{\cal D}^j(\hat U)_{mm'} \; \sqrt{j(j+1) - m'(m'-1)} \;
\Phi_{jm'-1}(b_1,b_2)\cr
}$$
By comparing the expressions for $L_+\Psi$ and $\hat L_- \Psi$,
we deduce
$$\hat L_- \Psi = L_+ \Psi = L_-^\dagger \Psi$$
which completes the proof. Similarly, we would obtain
$$\hat L_+ \Psi = L_-\Psi\qquad \hat L_z \Psi = L_z \Psi\qquad \hat K_\pm \Psi
= K_\mp \Psi\qquad \hat K_z \Psi = K_z \Psi$$
so that the images $\widehat {\vec L}$ and $\widehat {\vec K}$ of $\vec L$
and $\vec K$ satisfy (21).
\smallskip
Equation (21) can be generalized in the
form $A^\dagger \Psi = \hat A\Psi$. The latter
relation allows to derive the image $\hat A$ of $A$
(or the operator $A$ if we know $\hat A$) [29,30].

\vfill\eject

\centerline {\bf Appendix 3:
Laplacians for the KS transformation}
\smallskip
\bigskip
This appendix is devoted to the volume elements $dv$, the line
elements $ds^2$ and the Laplacians $\Delta$ in {\msim R}$^3$ and {\msim
R}$^4$. It covers results obtained by Kustaanheimo and Stiefel
[1], Ikeda and Miyachi [4], Boiteux [5], Kibler and N\'egadi
[8], Polubarinov [10], D'Hoker and Vinet [32], and Lambert and Kibler
[16] for the KS transformation. We
shall deal in this appendix with the KS transformation in the form
corresponding to the {\msim R}$^4 \to \hbox {\msim R}^3$ map defined by
equations (4) and (14-16).
\smallskip
{\bf Volume elements.} The volume elements in {\msim R}$^4$ and
{\msim R}$^3$ are given by
$$
dv(\hbox{\msim R}^4) = du_1\,du_2\,du_3\,du_4 = {1\over 16}
\; r \; \sin \theta \; dr \; d\psi \; d\theta \; d\varphi
$$
$$
dv(\hbox {\msim R}^3) = dx_1\,dx_2\,dx_3 = r^2 \; \sin \theta
\; dr \; d\theta \; d\varphi
$$
Thus, we have
$$
dv(\hbox {\msim R}^4) = {1\over 16r} \; dv (\hbox {\msim R}^3) \; d\psi
$$
Integration on $\psi$ from 0 to $4\pi$ (i.e., integration on the fiber of the
{\msim R}$^4 - \{0\}
=S^3 \times \hbox {\msim R}^{+ \ast} \to {\msim R}^3 - \{0\} =
S^2 \times {\msim R}^{+ \ast}$ transformation) leads to
$$
\int_{\hbox {\msim R}^4} \ldots dv (\hbox {\msim R}^4) = {\pi\over 4}\,
\int_{\hbox {\msim R}^3} \ldots {1\over r} \, dv (\hbox {\msim R}^3)
$$
The latter equation allows us to formally define a Jacobian for the KS
transformation. Indeed, we have
$$dv(\hbox {\msim R}^3) = \vert J \vert \,  dv(\hbox {\msim R}^4)$$
where
$$ J = \hbox {det}
\pmatrix{
{\partial x_1\over\partial u_1}&{\partial x_1\over\partial u_2}&
{\partial x_1\over\partial u_3}&{\partial x_1\over\partial u_4}\cr
& & & \cr
{\partial x_2\over\partial u_1}&{\partial x_2\over\partial u_2}&
{\partial x_2\over\partial u_3}&{\partial x_2\over\partial u_4}\cr
& & & \cr
{\partial x_3\over\partial u_1}&{\partial x_3\over\partial u_2}&
{\partial x_3\over\partial u_3}&{\partial x_3\over\partial u_4}\cr
& & & \cr
{1\over 8\pi u_2}&{-1\over 8\pi u_1}&{1\over 8\pi u_4}&{-1\over 8\pi u_3}\cr
}=
{4\over \pi} \> (u^2_1+u^2_2+u^2_3+u^2_4)$$
plays the r\^ole of a Jacobian (the origin of which is not
clear) for the considered {\msim R}$^4 \to $ {\msim R}$^3$ map.
\smallskip
{\bf Line elements.} The line element of {\msim R}$^4$
$$
ds^2(\hbox{\msim R}^4) = du^2_1 + du^2_2 + du^2_3 + du^2_4 =
dz_1d\bar z_1 + dz_2d\bar z_2
$$
is given in the coordinates $(\sqrt{r},\psi, \theta, \varphi)$ by
$$
ds^2(\hbox{\msim R}^4) = {1\over4r} \; [dr^2 + r^2 \; d\theta^2
+ r^2(d\psi^2 + 2 \; \cos \theta \; d\psi \; d\varphi + d\varphi ^2)]
$$
Similarly, the line element of {\msim R}$^3$
$$
ds^2(\hbox{\msim R}^3) = dx^2_1 + dx_2^2 + dx^2_3
$$
is known to be
$$
ds^2(\hbox{\msim R}^3) = dr^2 + r^2 \, d \theta ^2 + r^2 \,
\sin ^2 \theta \; d\varphi ^2
$$
in the coordinates $(r,\theta,\varphi)$. Therefore, the connection between
$ds^2(\hbox{\msim R}^3)$ and $ds^2(\hbox{\msim R}^4)$ is
$$
4r\,ds^2(\hbox{\msim R}^4) = ds^2(\hbox{\msim R}^3)+r^2
(\cos \theta \; d\varphi + d\psi)^2
$$
The constraint
$$
\cos\theta \; d\varphi + d\psi = 0
$$
ensures that
$$
ds^2(\hbox{\msim R}^3) = 4r\,ds^2(\hbox{\msim R}^4)
$$
In other words, the $\hbox {\msim R}^4 \to \hbox {\msim R}^3$ map defined by
equations (16) is conformal once the one-form $\omega$ given by
$$
\omega = 2(u_2du_1 - u_1du_2 + u_4du_3 - u_3du_4) =
- r \, (\cos \theta \; d\varphi + d\psi)
$$
is taken to vanish (see Kustaanheimo and Stiefel [1]). However, the relation
$$
ds^2(\hbox{\msim R}^3) = 4r\,ds^2(\hbox{\msim R}^4) - \omega^2
$$
is to be used if $\omega \ne 0.$
\smallskip
{\bf Laplacians.} The Laplace-Beltrami operator
$$
\Delta = {1\over \sqrt{\Vert G \Vert}}\; {\partial\over{\partial\xi^a}}
\; {\sqrt{\Vert G \Vert}} \; G^{ab} \, {\partial\over
{\partial\xi^b}}\qquad
\Vert G \Vert = | \det (G_{ab}) |
$$
where the metric matrix $(G^{ab}) = (G_{ab})^{-1}$ is defined via
$$
ds^2 = G_{ab} \, d\xi^a \, d\xi^b
$$
can be particularized to $\hbox {\msim R}^4$ and $\hbox {\msim R}^3$ as
follows.
In the Euler-angle coordinates $({\sqrt r}, \psi, \theta, \varphi)$ of
$\hbox {\msim R}^4$, we get the Laplacian
$$
\Delta (\hbox{\msim R}^4) = {4 \over r} \, {\partial \over {\partial r}}
\, \left( r^2 {\partial \over {\partial r}} \right) +
{4 \over {r \sin \theta}} \, {\partial \over {\partial\theta}} \, \left(
\sin\theta {\partial \over \partial \theta} \right)
+ {4 \over {r \sin^2 \theta}} \,
\left( {{\partial^2} \over {\partial \varphi^2}}
     + {{\partial^2} \over {\partial \psi^2}}
\right) - {{8 \cos \theta} \over {r \sin^2 \theta}} \,
{\partial^2 \over {\partial \psi \partial \varphi}}
$$
while in the spherical coordinates $(r, \theta, \varphi)$ of $\hbox{\msim R}^3$
we have the usual Laplacian
$$
\Delta (\hbox {\msim R}^3) = {1\over {r^2}}\, {\partial\over
{\partial r}} (r^2 {\partial\over{\partial r}}) +
{1\over{r^2 \sin \theta}}\, {\partial\over {\partial\theta}}\,
(\sin \theta {\partial \over {\partial\theta}}) +
{1 \over {r^2 \sin^2 \theta}}\, {\partial^2 \over {\partial\varphi^2}}
$$
Consequently, the two Laplacians are connected through
$$
{1\over 4r} \, \Delta (\hbox {\msim R}^4) =
\Delta (\hbox {\msim R}^3) +
{1\over {r^2 \sin^2 \theta}}\, {\partial \over {\partial \psi}}
({\partial \over {\partial \psi}}
- 2 \cos \theta {\partial\over {\partial \varphi}})
$$
When acting on (two-fold differentiable) functions of the variables $x_1, x_2$
and $x_3$, the latter relation reduces to
$$
\Delta (\hbox {\msim R}^3) =
{1\over 4}\; {1\over {u^2_1 + u^2_2 + u^2_3 + u^2_4}}\;
\Delta (\hbox {\msim R}^4)
$$
in agreement with a well-known result
(see Boiteux [5] and  Kibler and N\'egadi [8]).
However, when $\Delta (\hbox {\msim R}^3)$ acts on (two-fold differentiable)
functions of the variables $u_1,u_2,u_3$ and $u_4$, we have
$$
\Delta (\hbox{\msim R}^3) = {1\over {4r}}\, \Delta (\hbox{\msim R}^4) +
{1\over {r^2}}\, K^2_3 +
{{\cos \theta}\over {r^2 \sin^2 \theta}}\; K_3 (\cos \theta \> K_3 - 2L_3)
$$
The third term in the right-hand side of the latter general
formula does not occur in the corresponding formula of Ref.~[8] ; the
geometrical interpretation of this fact is an interesting
problem, especially in the case of a general Hurwitz
transformation. Finally, note that by introducing
$$
\lambda = {1 \over 2} (r + x_3) \qquad
    \mu = {1 \over 2} (r - x_3) \qquad
      A = - i(K_3 + L_3) \qquad
      B = - i(K_3 - L_3)
$$
we have
$$
{1\over {r^2}}\, K^2_3 +
{{\cos \theta}\over {r^2 \sin^2 \theta}}\; K_3 (\cos \theta\, K_3 - 2L_3)
= - {1 \over{16}} {1 \over{\lambda + \mu}} {1 \over{\lambda \mu}}
[(- \lambda + 3 \mu) A^2 + (3 \lambda - \mu)B^2
+ 2(\lambda + \mu)AB]
$$
so that the difference $\Delta (\hbox{\msim R}^3) - (4r)^{-1} \Delta
(\hbox{\msim R}^4)$ can be rewritten (in the general case) in a somewhat
symmetric form.
\smallskip
We close this appendix with a remark concerning the canonical
character of the KS transformation. In fact, we can verify that
$$
\sum_{i = 1}^{3}      dx_i        \; {\partial \over {\partial x_i}} =
\sum_{\alpha = 1}^{4} du_{\alpha} \; {\partial \over {\partial u_{\alpha}}}
$$
provided that the constraint
$$
  {1 \over {2r}} \, \omega \; X \equiv
- {i \over {r}}  \, \omega \; {K}_3 = 0
$$
be satisfied. The canonicity of the KS transformation arises
from the latter two relations (to be compared with
$\sum \, p_k \, dq_k = \sum \, P_k \, dQ_k + dF$
characterizing a canonical or contact transformation in
classical mechanics).
\smallskip
{\bf Extensions.} The results contained in this appendix can be
developed for the non-compact extensions of the $\grq \to \grt$
map defined by (4) and (14-16).  These extensions are of the
type $\gr^{2,2} \to \gr^{1,2}$ or $\gr^{2,2} \to \gr^{2,1}$
and are associated to fibrations on hyperboloids with compact
or non-compact fiber, respectively (see Ref.~[16]).

\vfill\eject

\centerline {\bf Appendix 4: Differential aspects
of the $\grh \to \grc$ transformation}
\smallskip
\bigskip
We shall deal in this appendix with the transformation defined through
equations (55) and (58). As a preliminary, we note that by taking
$$
u_3 = u_4 = u_7 = u_8 = 0
$$
and by making the replacements
$$
u_1 \to u_2 \quad
u_5 \to u_4 \quad
u_2 \to u_1 \quad
u_6 \to u_3
$$
the $\grh \to \grc$ map defined by
(55) reduces to the $\grq \to \grt$ map defined by (16). Therefore, we may
think to introduce
in an easy way Euler-angle coordinates $(\sqrt r, \psi, \theta,
\varphi, \theta_1, \theta_2, \theta_3, \theta_4)$ for the $\grh \to \grc$
transformation.
\smallskip
{\bf Euler-angle coordinates.} It is sufficient to work with
the inverse transformation, defined up to an $S^3$ sphere,
for obtaining the $u_\alpha$'s in terms of, what we call,
the Euler-angle coordinates $(\sqrt r, \psi, \theta,
\varphi, \theta_1, \theta_2, \theta_3, \theta_4)$. In fact, we have
$$
\eqalign{
u_1 & = \sqrt r \, \cos {{\theta_1} \over 2} \,
\cos \psi \, \cos \theta \, \cos \varphi\cr
u_2 & = \sqrt r \, \cos {{\theta_1} \over 2} \,
\sin \psi \, \cos \theta \, \cos \varphi\cr
u_3 & = \sqrt r \, \cos {{\theta_1} \over 2} \,
\sin \theta \, \cos \varphi\cr
u_4 & = \sqrt r \, \cos {{\theta_1} \over 2} \,
\sin \varphi\cr
u_5 & = \sqrt r \, \sin {{\theta_1} \over 2} \, (
- \sin \theta_2 \, \sin \theta_3 \, \cos \theta_4 \, \sin \theta \, \cos
\varphi
- \sin \theta_2 \, \cos \theta_3 \, \sin \psi \, \cos \theta \, \cos \varphi\cr
    & + \cos \theta_2 \, \cos \psi \, \cos \theta \, \cos \varphi
- \sin \theta_2 \, \sin \theta_3 \, \sin \theta_4 \, \sin \varphi)\cr
u_6 & = \sqrt r \, \sin {{\theta_1} \over 2} \, (
- \sin \theta_2 \, \sin \theta_3 \, \cos \theta_4 \, \sin \varphi
+ \sin \theta_2 \, \cos \theta_3 \, \cos \psi \, \cos \theta \, \cos \varphi\cr
    & + \cos \theta_2 \, \sin \psi \, \cos \theta \, \cos \varphi
+ \sin \theta_2 \, \sin \theta_3 \, \sin \theta_4 \,
\sin \theta \, \cos \varphi)\cr
u_7 & = \sqrt r \, \sin {{\theta_1} \over 2} \, (
+ \sin \theta_2 \, \sin \theta_3 \, \cos \theta_4 \,
\cos \psi \, \cos \theta \, \cos \varphi\,
+ \sin \theta_2 \, \cos \theta_3 \, \sin \varphi\cr
    & + \cos \theta_2 \, \sin \theta \, \cos \varphi \,
- \sin \theta_2 \, \sin \theta_3 \, \sin \theta_4 \,
\sin \psi \, \cos \theta \, \cos \varphi)\cr
u_8 & = \sqrt r \, \sin {{\theta_1} \over 2} \, (
+ \sin \theta_2 \, \sin \theta_3 \, \cos \theta_4 \,
\sin \psi \, \cos \theta \, \cos \varphi \,
- \sin \theta_2 \, \cos \theta_3 \, \sin \theta \, \cos \varphi\cr
    & + \cos \theta_2 \, \sin \varphi \,
+ \sin \theta_2 \, \sin \theta_3 \, \sin \theta_4 \,
\cos \psi \, \cos \theta \, \cos \varphi)\cr
}
$$
Then, by introducing the latter equations into (55), we obtain
$$
\eqalign{
x_5 & = r \cos \theta_1 \cr
x_1 & = r \sin \theta_1 \cos \theta_2 \cr
x_2 & = r \sin \theta_1 \sin \theta_2 \cos \theta_3 \cr
x_3 & = r \sin \theta_1 \sin \theta_2 \sin \theta_3 \cos \theta_4 \cr
x_4 & = r \sin \theta_1 \sin \theta_2 \sin \theta_3 \sin \theta_4 \cr
}
$$
and $(r, \theta_1, \theta_2, \theta_3, \theta_4)$ are clearly the ``spherical''
coordinates for $\grc$.
\smallskip
{\bf Line elements.} From equation (55), we easily check that the line elements
$$
ds^2(\grh) = \sum_{\alpha=1}^8 \, du_\alpha^2 \qquad
ds^2(\grc) = \sum_{i=1}^5 \, dx_i^2
$$
are connected through
$$
ds^2(\grc) = 4 r \, ds^2(\grh) - (\omega_1^2 + \omega^2_2 + \omega^2_3)
$$
where the one-forms $\omega_i$ ($i = 1,2,3$) are defined in (59).
\smallskip
{\bf Laplacians.} The Laplacians $\Delta(\grh)$ and $\Delta(\grc)$
can be expressed in the coordinates $(\sqrt r, \psi, \theta, \varphi,
\theta_1, \theta_2, \theta_3, \theta_4)$. For saving space purposes,
the relation between $\Delta(\grh)$ and $\Delta(\grc)$ is not given
here. Note, however, that this relation reduces to
$$
\Delta(\grc) =
\left(4 \sum_{\alpha=1}^8\, u_\alpha^2\right)^{-1} \;
\Delta(\grh)
$$
when acting on a (two-fold) differentiable function
$f(x_1, x_2, \ldots, x_5)$. The corresponding relation to be used when
$f(x_1, x_2, \ldots, x_5)$ is replaced by a (two-fold) differentiable function
$g(u_1, u_2, \ldots, u_8)$ can be obtained from one of the authors (M.~K.).
\smallskip
Let us close this article by noticing that the $\grh \to \grc$ transformation
(55-59) turns out to be a canonical transformation since we have
$$
\sum_{i = 1}^{5}      dx_i        \; {\partial \over {\partial x_i}} =
\sum_{\alpha = 1}^{8} du_{\alpha} \; {\partial \over {\partial u_{\alpha}}}
$$
once the constraint
$$
  {1 \over {2r}} \, \sum_{j = 1}^{3} \, \omega_j \; X_j \equiv
- {i \over {r}}  \, \sum_{j = 1}^{3} \, \omega_j \; {\cal K}_j = 0
$$
is satisfied.
\smallskip
{\bf Extensions.} The results contained in this appendix can be
developed for the non-compact extensions of the $\grh \to \grc$
map defined by (55) and (58).  These extensions are of the type
$\gr^{4,4} \to \gr^{1,4}$ or $\gr^{4,4} \to \gr^{3,2}$ and are
associated to fibrations on hyperboloids with compact or
non-compact fiber, respectively (see Ref.~[16]).

\vfill\eject

\centerline {\bf References}
\parskip = 0 true cm
\smallskip
\bigskip
\baselineskip = 0.72 true cm
\item {[\ 1]} Kustaanheimo, P. and Stiefel, E.,
J. reine angew. Math. {\bf 218}, 204 (1965).

\item {[\ 2]} Hopf, H., Math. Ann. {\bf 104}, 637 (1931). See
also: Gluck, H. and Warner, F.W., Duke Math. J. {\bf 50}, 107 (1983).

\item {[\ 3]} Souriau, J.-M., {\it Structure des syst\`emes
dynamiques}, Dunod: Paris (1970).

\item {[\ 4]} Ikeda, M. and Miyachi, Y.,
Math. Japon. {\bf 15}, 127 (1970).

\item {[\ 5]} Boiteux, M.,
C.R. Acad. Sci. (Paris), S\'er. B, {\bf 274}, 867 (1972);
Physica {\bf 75}, 603 (1974);
J. Math. Phys. {\bf 23}, 1311 (1982).

\item {[\ 6]} Hage Hassan, M., Grenet, G., Gazeau, J.-P. and Kibler,
M., J. Phys. A: Math. Gen. {\bf 13}, 2623 (1980).

\item {[\ 7]} Blanchard, Ph. and Sirugue, M.,
J. Math. Phys. {\bf 22}, 1372 (1981).

\item {[\ 8]} Kibler, M. and N\'egadi, T.,
Lett. Nuovo Cimento {\bf 37}, 225 (1983);
J. Phys. A: Math. Gen. {\bf 16}, 4265 (1983);
Phys. Rev. A {\bf 29}, 2891 (1984).

\item {[\ 9]} Vivarelli, M.D.,
Celest. Mech. {\bf 29}, 45 (1983); {\bf 36}, 349 (1985);
{\bf 41}, 359 (1988); ``The KS-Transformation Revisited'',
Report n.26/p, Politecnico di Milano, Milano (1991).

\item {[10]} Polubarinov, I.V., ``On the Application of Hopf Fiber
Bundles in Quantum Theory'', Preprint E2-84-607, JINR: Dubna (1984).

\item {[11]} Cornish, F.H.J.,
J. Phys. A: Math. Gen. {\bf 17}, 2191 (1984).

\item {[12]} Kibler, M. and N\'egadi, T.,
Croat. Chem. Acta {\bf 57}, 1509 (1984).

\item {[13]} Iwai, T.,
J. Math. Phys. {\bf 26}, 885 (1985); Iwai, T. and Rew, S.-G.,
Phys. Lett. {\bf 112A}, 6 (1985).

\item {[14]} Kibler, M., Ronveaux, A. and N\'egadi, T.,
J. Math. Phys. {\bf 27}, 1541 (1986).

\item {[15]} Mardoyan, L.G., Pogosyan, G.S., Sissakian,
A.N. and Ter-Antonyan, V.M., ``Exact Solution to the
Kibler-Ronveaux-N\'egadi Problem'',
Preprint P2-86-431, JINR: Dubna (1986).

\item {[16]} Lambert, D., Kibler, M. and Ronveaux, A., ``On
Nonbijective Canonical Transformations in Mathematical
Physics'', in {\it Group Theoretical Methods in Physics},
ed. Y.M. Cho, World Scientific: Singapore (1986).
Lambert, D. and Kibler, M., ``Levi-Civita, Kustaanheimo-Stiefel
and Other Transformations'', in {\it Group Theoretical Methods in Physics},
ed. R. Gilmore, World Scientific: Singapore (1987).
Lambert, D. and Kibler, M.,
J. Phys. A: Math. Gen. {\bf 21}, 307 (1988).

\item {[17]} Kibler, M., ``Non-bijective Canonical
Transformations'', Report LYCEN 8713, IPN: Lyon (1987).

\item {[18]} Davtyan, L.S., Mardoyan, L.G., Pogosyan, G.S., Sissakian,
A.N. and Ter-Antonyan, V.M.,
J. Phys. A: Math. Gen. {\bf 20}, 6121 (1987).

\item {[19]} Kibler, M. and Winternitz, P.,
J. Phys. A: Math. Gen. {\bf 21}, 1787 (1988).

\item {[20]} Kibler, M., ``Application of Non-bijective
Transformations to Various Potentials'',
in {\it Group Theoretical Methods in Physics},
eds. H.-D. Doebner, J.-D. Hennig and T.D. Palev, Springer: Berlin (1988).

\item {[21]} Cordani, B., ``On the Generalization of
Kustaanheimo-Stiefel Transformation'',
Report Quaderno n.~49, Universit\`a di Milano, Milano (1988).

\item {[22]} Kibler, M. and Labastie, P., ``Transformations
Generalizing the Levi-Civita, Kusta\-an\-heimo~-~Stiefel, and Fock
Transformations'', in {\it Group Theoretical Methods in Physics},
eds. Y. Saint-Aubin and L. Vinet, World Scientific: Singapore (1989).

\item {[23]} Hage Hassan, M. and Kibler, M., ``Non-bijective
Quadratic Transformations and Theory of Angular Momentum'',
in {\it Selected Topics in Statistical Mechanics}, eds. A.A.
Logunov, N.N. Bogolubov, Jr., V.G. Kadyshevsky and A.S.
Shumovsky, World Scientific: Singapore (1990).

\item {[24]} Fock, V.A., Z. Phys. {\bf 98}, 145 (1935).

\item {[25]} Fock, V.A., Z. Phys. {\bf 49}, 339 (1928).

\item {[26]} Bargmann, V., Rev. Mod. Phys. {\bf 34}, 829 (1962).

\item {[27]} Schwinger, J., in {\it Quantum Theory of Angular Momentum},
eds. L.C. Biedenharn and H. van Dam, Academic: New York (1965).

\item {[28]} Woronowicz, S.L., Commun. Math. Phys. {\bf 111}, 613 (1987).

\item {[29]} Hage Hassan, M., J. Phys. A: Math. Gen. {\bf 12}, 1633
(1979); ``Th\`ese d'Etat'', Universit\'e Claude Bernard Lyon-1 (1980).

\item {[30]} Gulshani, P., Can. J. Phys. {\bf 57}, 998 (1979).

\item {[31]} Talman, J.D., {\it Special Functions. A Group Theoretic
Approach}, Based on Lectures by E.P. Wigner, Benjamin: New York (1968).

\item {[32]} D'Hoker, E. and Vinet L., Lett. Math. Phys. {\bf 12}, 71 (1986).

\bye